\documentclass[twocolumn,aps,pra,showpacs]{revtex4-1}

\usepackage{graphicx}
\usepackage{dcolumn}
\usepackage{bm}
\usepackage{amsmath}

\begin{document}

\title{Quantum description of electromagnetic fields in waveguides}

\author{Akira Kitagawa}
\email{kitagawa@kochi-u.ac.jp}

\affiliation{Faculty of Education, Kochi University, 2-5-1 Akebono-cho, Kochi 780-8520, Japan}

\begin{abstract}
Using quantum theory, we study the propagation of an optical field in an inhomogeneous dielectric, 
and apply this scheme to traveling optical fields in a waveguide. 
We introduce a field-atom interaction Hamiltonian and derive the refractive index 
using quantum optics. We show that the transmission and reflection 
of optical fields at an interface between different materials can be described 
with normalized Fresnel coefficients and that this representation is related 
to the beam splitter operator. We then study the propagation properties of the optical 
fields for two types of slab waveguides: step-index and graded-index. The waveguides 
are divided into multiple layers to represent the spatial dependence of the optical field. 
We can evaluate the number of photons in an arbitrary volume in the waveguide 
using this procedure. 
Using the present method, the quantum properties of weak optical fields in a waveguide 
are revealed, while coherent states with higher amplitudes 
reduces to representation of classical waveguide optics.  
\end{abstract}


\maketitle 

\section{Introduction}
Currently, optical technology plays an important role in communication. 
To achieve long-distance communication, signals are transmitted 
through optical waveguides, which typically consist of silica-based glass.

The optical waveguide is a device used to achieve free propagation 
along its long axis and a confinement over its cross-section. 
Conventional optical waveguides consist of a core and cladding 
and can confine optical fields in the core 
because of the difference in indices. These optical waveguides are classified 
into two types: step-index (SI) and graded-index (GI). 
In particular, a waveguide of concentric structure is referred to as 
an optical fiber.

Conventionally, it is assumed that optical signals 
exhibit a certain amount of intensity in a waveguide, and 
the mathematical structure of the SI and GI waveguides 
has been very well studied within classical waveguide optics \cite{Snitzer61,Snyder_text}. 
In addition, protocols using quantum interference among weak optical signals, 
referred to as quantum protocols, have recently been proposed 
\cite{Nielsen_text,Braunstein_text}. 
In these protocols, optical signals should be described using quantum optics. 
The optical waveguide is also considered an important device 
for long-distance propagation in quantum protocols, 
such as in quantum cryptography \cite{Gisin07, Sasaki11}.

Optical fields in a waveguide interact with dielectrics, 
and the interaction causes an effective decrease in light speed; 
this effect is phenomenologically introduced by a refractive index 
that is typically greater than or equal to unity. This process is 
discussed semi-classically in \cite{Feynmann_text,Hecht_text}. 
Thus far, however, the quantization of optical waves traveling 
in a waveguide has not been discussed adequately. 
One of the few exceptions is the study in \cite{Drummond01}. 
In this study, boundary effects in the waveguide were neglected, 
implying that the optical wave was assumed to travel along the core in a straight line. 
This assumption corresponds to a single-mode waveguide case. 
Strictly speaking, however, the optical wave in a waveguide 
shows spatial distribution over the cross-section, and 
the propagation property of the optical field is generally determined 
by the boundary condition at the interface between the core and cladding. 
In this sense, the spatial dependence of the optical field should also be considered 
when studying optical waveguides within quantum optics.

In this research, we study the propagation of the optical field in a dielectric 
such as silica-based glass from a quantum optics perspective 
and assume that the optical field is in a coherent state to allow 
comparison with the classical waveguide optics. 
First, we treat both the optical field and atom quantum theoretically 
and introduce an interaction Hamiltonian between them. 
Using quantum optics, we show that the refractive index can consistently be described 
as in a coherent state and that the electromagnetic field can be represented 
by this refractive index.

We also study transmission and reflection of optical fields 
at an interface between different materials. These properties 
are described by Fresnel coefficients in classical optics \cite{Hecht_text}, 
where, to satisfy the energy conservation law, the transmission coefficient 
is corrected by a factor that include refractive indices \cite{Hecht73, Zia88}. 
Here as an alternative, 
we normalize electromagnetic fields and introduce normalized Fresnel 
coefficients. We show that this scheme is consistent 
with the representation of the beam splitter operator \cite{Barnett_text}.

Following this, we discuss the propagation properties of optical fields 
in optical waveguides for SI and GI types. 
In both cases, the waveguides are divided into multiple layers, 
and the electromagnetic fields in each layer are described. 
This multi-layer division method enables us to represent the spatial dependence 
of the optical fields using quantum optics. 
Electromagnetic fields in adjacent layers are associated with each other 
using normalized Fresnel coefficients, 
and various propagation properties of the waveguides are studied. 
Here, we focus our attention on the linear properties of optical waveguides. 
We describe the quantum properties of optical fields in a waveguide 
for weak coherent states, while coherent states with higher amplitudes 
are reduced to a description in classical waveguide optics.

The rest of this paper is organized as follows. In Sec. \ref{dielectrics}, 
we describe the propagation of the optical field in a dielectric. 
In Sec. \ref{Fresnel}, from a quantum optics perspective, 
we consider the transmission and reflection of the optical fields 
at an interface between different materials. In Sec. \ref{waveguide}, 
we study the propagation properties of the optical field 
in SI and GI optical waveguides. Section \ref{summary} presents the summary. 
Detailed derivations of the mathematical formulae used are presented in the appendices.

\section{Propagation of the optical field in a dielectric}
\label{dielectrics}
\subsection{Interaction-free Hamiltonian}

Let us start with the vector potential as follows: 
\begin{eqnarray}
\hat{\bm{A}}_l &=&\sqrt{\frac{\hbar }{2\omega \varepsilon _0 V} } 
\bigg\{ \hat{a}_l ^\dagger \exp [i(\omega t-\bm{k}_l \cdot \bm{r} )] \nonumber \\
&&\hspace{10mm} +\hat{a}_l \exp[-i(\omega t-\bm{k}_l \cdot \bm{r})]\bigg\} \bm{e}_l , 
\end{eqnarray}
where $\omega $ and $\varepsilon _0 $ are the frequency of the optical field 
and the permittivity in vacuum, respectively, and $V=L_x L_y L_z $ is a unit volume, which 
is given later. The operators $\hat{a}_l ^\dagger $ and $\hat{a}_l $ are the creation 
and annihilation operators in a mode $l$, respectively, and they satisfy the commutation 
relation 
\begin{equation}
[\hat{a}_l ,\hat{a}_{l' } ^\dagger ]=\delta _{ll'}. \label{commutation_relation}
\end{equation}
The vectors $\bm{k}_l $ and $\bm{e}_l $ are perpendicular to each other 
and are related to the wave vector and direction of polarization, respectively. 
For consistency with the later part, we employ $\bm{k}_l =\kappa _l \bm{e}_x +\beta _l \bm{e}_z $ 
and $\bm{e}_l =\bm{e}_y $. The magnitude of the vector $\bm{k}_l $ 
corresponds to the wave number in vacuum: 
\begin{equation}
|\bm{k}_l |=\sqrt{\kappa _l ^2 +\beta _l ^2 }=k_l . 
\end{equation}

Considering the Coulomb gauge, we can obtain the electromagnetic fields thus 
\begin{subequations}
\begin{eqnarray}
\hat{\bm{E}}_l &=&-\frac{\partial \hat{\bm{A}}_l }{\partial t} \nonumber \\
&=&-i\sqrt{\frac{\hbar \omega }{2\omega \varepsilon _0 V} } 
[U_{tz} ^\ast \hat{a}_l ^\dagger \exp (-i\kappa _l x) \nonumber \\
&&\hspace{15mm} -U_{tz} \hat{a}_l \exp (i\kappa _l x)]\bm{e}_y , 
\label{electric_field} \\
\hat{\bm{H}}_l &=&\mu _0 ^{-1} (\nabla \times \hat{\bm{A}}_l ) \nonumber \\
&=&-\frac{i}{\mu _0 } \sqrt{\frac{\hbar }{2\omega \varepsilon _0 V} } 
[U_{tz} ^\ast \hat{a}_l ^\dagger \exp (-i\kappa _l x) \nonumber \\
&&\hspace{15mm} -U_{tz} \hat{a}_l \exp(i\kappa _l x)](-\beta \bm{e}_x +\kappa _l \bm{e}_z ), 
\label{magnetic_field}
\end{eqnarray}
\end{subequations}
where 
\begin{equation}
U_{tz} =\exp [-i(\omega t-\beta _l z)] \label{U_tz}
\end{equation}
is a temporal- and $z$-dependence. These electromagnetic fields propagate 
at light speed in vacuum along a direction 
parallel to $\bm{k}_l $, say $c$.

For the interaction-free case, equivalent to propagation in vacuum, 
the Hamiltonian is calculated using Eqs. (\ref{commutation_relation}), 
(\ref{electric_field}), and (\ref{magnetic_field}) as follows: 
\begin{eqnarray}
\hat{\mathcal{H}}_l &=&\int _V \left( \frac{\varepsilon _0 }{2} \hat{\bm{E}}_l ^2 
+\frac{\mu _0 }{2} \hat{\bm{H}}_l ^2 \right) dV \nonumber \\
&=&\hbar \omega \left( \hat{a}_l ^\dagger \hat{a}_l +\frac{1}{2} \right) 
\label{free_Hamiltonian_field}
\end{eqnarray}
Here, we can arbitrarily choose integration ranges along the $x$- and $y$-directions, 
for example, for $L_x =L_y =1$; a periodic boundary condition 
along the $z$-direction is considered. 
\begin{equation}
L_z =\frac{2\pi n_z }{\beta _l } \quad (n_z =1,2,\cdots ). 
\end{equation}

\subsection{Field-atom interaction Hamiltonian}

\begin{figure}[t]
\centering 
\includegraphics[bb=20 165 530 365, clip, width=.9\linewidth]{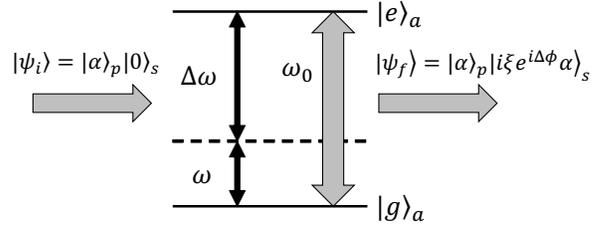}
\caption{\label{Fig1} Field-atom interaction in an off-resonant regime. 
The atom has two energy-levels, $|g\rangle _a $ and $|e\rangle _a $, 
separated by frequency $\omega _0 $. The atom is initially 
in the lower energy state $|g\rangle _a $, 
then the primary coherent field with frequency $\omega (\ll \omega _0 )$ comes in. 
The atom is virtually excited for a short time then soon de-excited. 
Via this interaction a secondary optical field, with a phase-shift of almost 
$\pi /2$, is generated in the same direction as the primary field. }
\end{figure}

Following on, we study the case where the primary optical field interacts with an atom 
in a dielectric. The schematic is shown in Fig. \ref{Fig1}. 
As a result of this interaction, a secondary optical field 
is generated in the same direction as the primary field \cite{Hecht_text}. 
The modes of the primary and secondary fields are distinguished 
with indices $l=p, s$, respectively, and we have 
\begin{equation}
\bm{k}_p =\bm{k}_s =\kappa \bm{e}_x +\beta \bm{e}_z . 
\end{equation}
We assume that the atom has two discrete states $|g\rangle _a $ and $|e\rangle _a $ 
separated by a frequency $\omega _0 $, and that the atom is initially 
in the lower state $|g\rangle _a $. The free Hamiltonian of the optical fields is represented by 
Eq. (\ref{free_Hamiltonian_field}) with $l=p,s$ and that of atom is given by 
\cite{Knight_text} 
\begin{equation}
\hat{\mathcal{H}}_a =\frac{\hbar \omega _0 }{2}\hat{\sigma }_z , 
\end{equation}
where $\hat{\sigma }_z $ is a component of the Pauli spin-$\frac{1}{2} $ 
operator satisfying 
\begin{subequations}
\begin{eqnarray}
\hat{\sigma }_z |e\rangle _a &=&|e\rangle _a , \label{sigma_z|e>} \\
\hat{\sigma }_z |g\rangle _a &=&-|g\rangle _a , \label{sigma_z|g>}
\end{eqnarray}
\end{subequations}
and let the total free Hamiltonian be 
\begin{equation}
\hat{\mathcal{H}}_0 =\hat{\mathcal{H}}_p +\hat{\mathcal{H}}_s +\hat{\mathcal{H}}_a . 
\end{equation}
We also assume that 
\begin{equation}
\Delta \omega \equiv \omega _0 -\omega \gg 0. \label{Delta_omega}
\end{equation}
In this off-resonant situation the atom can be virtually excited in a short time permitted 
by the uncertainty relation, 
\begin{equation}
\Delta \tau \sim \frac{1}{\Delta \omega } . \label{Delta_tau}
\end{equation}
The atom is quickly de-excited, and it reverts to the $|g\rangle _a $ state. 
We assume that the primary optical field is in a coherent state 
$|\alpha \rangle _p $, and the initial state can be described as 
\begin{equation}
|\psi _i \rangle =|\alpha \rangle _p |0\rangle _s , \label{initial_state}
\end{equation}
where the atomic state is not displayed because it remains at $|g\rangle _a $ 
before and after interaction.

Hereafter, we employ an approximation rule throughout this paper: 
small terms of the first order are included, 
while those of the second or higher orders are ignored. 
Regarding the interaction as a perturbation, 
we can write the effective interaction Hamiltonian 
in the interaction picture as follows (see Appendix \ref{derivation_effective_Hamiltonian}): 
\begin{equation}
\hat{\mathcal{H}}_{\rm eff} \simeq -\hbar \chi (\hat{a}_p \hat{a}_s ^\dagger 
+\hat{a}_p ^\dagger \hat{a}_s ), \label{effective_Hamiltonian}
\end{equation}
where $\chi $ is proportional to the square of the field-atom coupling constant $g$ 
given in Eq. (\ref{coupling_constant}), 
and small terms of second or higher orders are not shown 
according to the approximation rule. 
The final state through the interaction is 
\begin{equation}
|\psi _f \rangle =\exp \left(-i\frac{\hat{\mathcal{H}}_{\rm eff} }{\hbar } 
\Delta \tau \right) 
|\psi _i \rangle 
=|\alpha \rangle _p |i\xi \exp (i\delta )\alpha \rangle _s , \label{final_state}
\end{equation}
where we used the relations in Appendix \ref{BS_op} 
and $\xi =\chi \Delta \tau$ is a small parameter. 
However, note that the leading term of the secondary mode is a small term 
of the first order, and that one of the second order
should be considered here. 
It appears as a small phase $\delta $ that should be kept, 
even in the present approximation rule. 
This small phase, however, has no effect 
on the result because it disappears by multiplication 
with another small parameter. 
We can also see from Eq. (\ref{final_state}) that the primary optical field 
is not attenuated at all within this approximation rule.

\subsection{Expectation value of the composite field}
The primary and secondary optical fields spatially overlap after the interaction. 
In the following discussion we study just the electric field 
but it is valid for both the electric and magnetic fields. 
We introduce the composite electric field as 
\begin{equation}
\hat{\bm{E}}_c =\hat{\bm{E}}_p +\hat{\bm{E}}_s , 
\label{composite_electric_field_operator}
\end{equation}
where the suffix $c$ means a composite field. 
We obtain the expectation values of the optical field after the interaction 
with Eq. (\ref{final_state}) 
\begin{equation}
\bm{E}_1 =\langle \psi _f |\hat{\bm{E}}_c |\psi _f \rangle 
=\bm{E}_p +\xi \bm{E}_s , \label{composite_electric_field}
\end{equation}
where 
\begin{subequations}
\begin{eqnarray}
\bm{E}_l &=&-i\sqrt{\frac{\hbar \omega }{2\varepsilon _0 V} } 
[U_{tz} ^\ast \alpha _l ^\ast \exp (-i\kappa x)\exp (-i\phi _l )  \nonumber \\
&&\hspace{15mm} -U_{tz} \alpha _l \exp (i\kappa x)\exp (i\phi _l )]\bm{e}_y , \\
&&\phi _l =\left\{ 
\begin{array}{cl}
0 & (l=p) \\
\frac{\pi }{2} +\delta & (l=s)
\end{array}
\right. .
\end{eqnarray}
\end{subequations}
Here the suffix 1 in the left-hand side of Eq. (\ref{composite_electric_field}) means 
that the interaction with an atom occurs once, and we also considered the relations 
\cite{Louisell_text} as follows: 
\begin{subequations}
\begin{eqnarray}
&&\exp (i\theta \hat{a}_l ^\dagger \hat{a}_l )
|\alpha \rangle _l =|\alpha \exp (i\theta )\rangle _l , \\
&&\exp (-i\theta \hat{a}_l ^\dagger \hat{a}_l )\hat{a}_l 
\exp (i\theta \hat{a}_l ^\dagger \hat{a}_l )=\hat{a}_l \exp (i\theta ) 
\end{eqnarray}
\end{subequations}
for an arbitrary real number $\theta $. 
In practice, both $\bm{E}_p $ and $\bm{E}_s $ propagate 
at light speed in vacuum, and have a common spatial period. However, 
the phase of the secondary field is delayed by almost $\pi /2$ compared with 
that of the primary. Therefore, these two fields 
destructively interfere with each other 
and a composite field, $\bm{E}_1 $ with a spatial period identical to $\bm{E}_p $ 
and $\bm{E}_s $, is generated. 
Let the composite electric field be 
\begin{eqnarray}
\bm{E}_1 &=&-i\mathcal{A}\sqrt{\frac{\hbar \omega }{2\varepsilon _0 V} } 
[U_{tz} ^\ast \alpha ^\ast \exp (-i\kappa x)\exp (-i\delta \phi )  \nonumber \\
&&\hspace{15mm} -U_{tz} \alpha \exp (i\kappa x)\exp (i\delta \phi )]\bm{e}_y 
\label{composite_electric_field_2}
\end{eqnarray}
with a constant $\mathcal{A}$ and a small phase $\delta \phi $.

Comparing Eqs. (\ref{composite_electric_field}) 
and (\ref{composite_electric_field_2}), we find $\mathcal{A}=1$ and 
$\delta \phi =\xi $ within the present approximation rule. This means 
that the amplitude of the composite field can be regarded as identical to the primary one 
and that the phase of the composite field is, 
when compared with $\bm{E}_p $, delayed by $\xi $ 
after the interaction with the atom. By substituting these parameters 
into Eq. (\ref{composite_electric_field_2}), 
we obtain 
\begin{eqnarray}
\bm{E}_1 &=&-i\sqrt{\frac{\hbar \omega }{2\varepsilon _0 V} } 
\{ \alpha ^\ast \exp (i\omega t)\exp [ -i(k_0 r+\xi )]
\nonumber \\
&&\hspace{10mm} 
-\alpha \exp (-i\omega t)\exp [i(k_0 r+\xi )]\} , 
\end{eqnarray}
where $r=x\cos \theta +z\sin \theta $ and $\tan \theta =\beta /\kappa $.

\begin{figure}[t]
\centering 
\includegraphics[bb=135 230 570 520, clip, width=.8\linewidth]{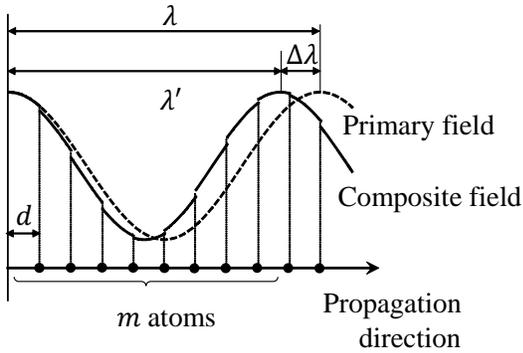}
\caption{\label{Fig2} Retardation in phase caused by field-atom interactions. 
The secondary optical field with phase shifted is generated 
via an interaction, and the composition of the primary and secondary fields 
results in a phase retardation. 
The optical field interacts successively with $m$ atoms within the spatial period 
of the primary field, and the wavelength of the composite field consequently 
becomes shorter than that of primary by $\Delta \lambda $. }
\end{figure}

The optical field repeatedly interacts with subsequent atoms in the dielectric 
and the phase is delayed with each interaction. We assume that the atoms 
are aligned in an orderly manner at a distance $d$ apart on the propagation path 
of the optical field. 
We also assume that the optical field interacts with $m$ atoms 
within its spatial period and that the phase 
is totally delayed by $m\xi $ (Fig. \ref{Fig2}). 
The total electric field is presented as 
\begin{eqnarray}
\bm{E}_m &=&-i\sqrt{\frac{\hbar \omega }{2\varepsilon _0 V} } 
\{ \alpha ^\ast \exp (i\omega t)\exp [ -i(k_0 r+m\xi )]
\nonumber \\
&&\hspace{10mm} 
-\alpha \exp (-i\omega t)\exp [i(k_0 r+m\xi )]\} . 
\end{eqnarray}
This successive retardation in phase makes the wavelength 
of the composite field, say, $\lambda ' $ shorter 
than that of the primary field $\lambda $ . 
For the difference in wavelength $\Delta \lambda =\lambda -\lambda ' $, we have 
\begin{equation}
k_0 \Delta \lambda =m\xi . 
\end{equation}
Since from the assumption we find $m=\lambda ' /d$, we can obtain 
\begin{equation}
\lambda '=\frac{\lambda }{n} , \label{wavelength}
\end{equation}
where 
\begin{equation}
n=1+\frac{\xi }{k_0 d} >1 \label{refractive_index}
\end{equation}
corresponds to the refractive index. From Eqs. (\ref{Delta_tau}), 
(\ref{coupling_constant}) and (\ref{chi}), 
and with the relation $k_0 =\omega /c$, we can see that 
$n-1\propto (\Delta \omega )^{-2} $. This means that $n$ is an increasing function 
with respect to $\omega $ over the range $\omega \ll \omega _0 $. This property agrees 
with Sellmeier's dispersion formula \cite{Born_text}.

From Eq. (\ref{wavelength}), the wave number in the dielectric accordingly becomes 
$k_0 ' =nk_0 $. Since the frequency $\omega $ stays unchanged during the interaction, 
we can regard the propagation speed of the composite field in the dielectric as 
$c' =c/n <c$.

So far, from the perspective of quantum optics, we have described 
the refractive index as a parameter $n$
that characterizes the interaction between the optical field and the dielectric. 
Henceforth, we study the propagation property 
of the optical field in a dielectric using this parameter.

\section{Quantum treatment of the transmission and reflection of the optical field}
\label{Fresnel}
In this section, we study the transmission and reflection of the optical field 
at an interface between different refractive indices from a quantum optics perspective.

\subsection{Normalization of the Fresnel coefficients}

\begin{figure}[t]
\centering 
\includegraphics[bb=45 150 485 460, clip, width=.9\linewidth]{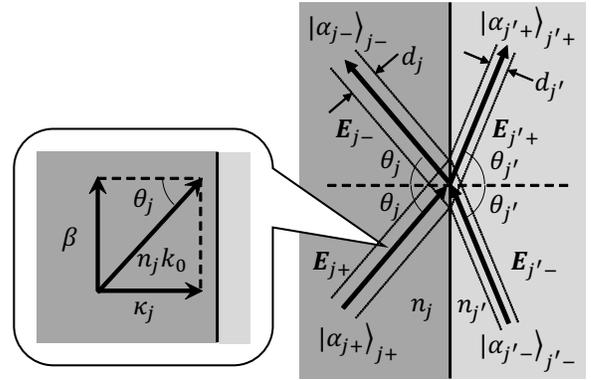}
\caption{\label{Fig3} Interactions among electric fields at an interface 
between different materials. Electric fields $j+$ and $j' -$ enter with 
incident angles $\theta _j $ and $\theta _{j' } $, respectively, 
and those of $j' +$ and $j-$ exit. Widths of the optical beams in the areas 
$j$ and $j' $ are denoted by $d_j $ and $d_{j' } $, respectively. }
\end{figure}

We consider an interface with refractive indices $n_j $ and $n_{j'} $. 
Two optical fields with indices $j+$ and $j' -$ enter this interface 
from opposite sides with incident angles of $\theta _j $ and $\theta _{j'} $, 
respectively, and interfere with each other. 
As a result, another two optical fields of indices $j- $ and $j' +$ 
exit the interface (Fig. \ref{Fig3}). Here '$j+$ ' and '$j-$' denote 
that the positive and negative $x$-component of the wave vector in an area $j$, respectively. 

We assume that the electric field is polarized in a direction parallel to the interface, 
which is the $y$-axis (TE mode), and that the wave vectors are given by 
$\bm{k}_{j\pm } =\kappa _j \bm{e}_x \pm \beta \bm{e}_z $ and 
$\bm{k}_{j' \pm } =\kappa _{j'} \bm{e}_x \pm \beta \bm{e}_z $, with double-sign in same order. 
Among the components of the wave vectors, the relation 
\begin{equation}
\kappa _i ^2 +\beta ^2 =(n_i k_0 )^2 
\end{equation}
holds for $i=j,j' $. 
Note that the $z$-component values are identical in all four optical fields 
according to Snell's law. 
The electromagnetic fields, that is, $\hat{\bm{E}}_{j\pm } $ and $\hat{\bm{H}}_{j\pm } $, 
are given with double-sign notation as follows: 
\begin{subequations}
\begin{eqnarray}
\hat{\bm{E}}_{j\pm } &=&-i\sqrt{\frac{\hbar \omega }{2\varepsilon _0 V} } 
[U_{tz} ^\ast \hat{a}_{j\pm } ^\dagger \exp (\mp i\kappa _j x) \nonumber \\
&&\hspace{2mm} -U_{tz} \hat{a}_{j\pm } \exp (\pm i\kappa _j x)]\bm{e}_y , \\
\hat{\bm{H}}_{j\pm } &=&-\frac{i}{\mu _0 } \sqrt{\frac{\hbar }{2\omega \varepsilon _0 V} } 
[U_{tz} ^\ast \hat{a}_{j\pm } ^\dagger \exp (\mp i\kappa _j x) \nonumber \\
&&\hspace{2mm} -U_{tz} \hat{a}_{j\pm } \exp(\pm i\kappa _j x)](-\beta \bm{e}_x 
\pm \kappa _j \bm{e}_z ), 
\end{eqnarray}
\end{subequations}
Classical electromagnetic fields correspond to the expectation values of the above operators 
when in a coherent state, that is, 
\begin{subequations}
\begin{eqnarray}
\bm{E}_{j\pm } &=&_{j\pm } \langle \alpha _{j\pm } |\hat{\bm{E}}_{j\pm } 
|\alpha _{j\pm } \rangle _{j\pm } , \\
\bm{H}_{j\pm } &=&_{j\pm } \langle \alpha _{j\pm } |\hat{\bm{H}}_{j\pm } 
|\alpha _{j\pm } \rangle _{j\pm } , 
\end{eqnarray}
\end{subequations}
where $\hat{a}_{j\pm } |\alpha _{j\pm } \rangle _{j\pm } 
=\alpha _{j\pm } |\alpha _{j\pm } \rangle _{j\pm } $ with double-sign notation.

In an ordinary procedure, tangential components of electromagnetic fields 
must be continuous at the interface \cite{Hecht_text}. 
We obtain Fresnel coefficients that are the ratios of the amplitude 
of the transmitted or reflected electromagnetic fields to the incident ones. 
In this conventional formulation, the energy density per unit time 
of the input and output optical fields are not conserved at the interface. 
This is because the following factors are not considered; 
first, as stated in the previous section, 
the ratio of the effective light-speed in areas $j$ and $j' $ is $n_j ^{-1} :n_{j'} ^{-1} $, 
and second, the cross-section of the transmitted optical field is different 
from that of the incident field due to refraction. 
The energy density is proportional to the inverse of the cross-section. 
With a unit length along the $y$-direction, thus the ratio of energy density 
is $d_j ^{-1} :d_{j' } ^{-1} =n_j /\kappa _j :n_{j' } /\kappa _{j' } $ 
because $d_i \propto \cos \theta _i =\kappa _i /(n_i k_0 )$ for $i=j,j' $. 
As a consequence, the ratio of the energy density 
in areas $j$ and $j'$ is $\kappa _j ^{-1} :\kappa _{j'} ^{-1} $ per unit time. 
To conserve energy density in the conventional procedure 
the transmission coefficient is multiplied 
by a correction factor \cite{Hecht73, Zia88}.

Here we use another technique; normalization of the electromagnetic field in area $j$ 
by a factor $\kappa _j ^{-1/2} $. 
\begin{subequations}
\label{continuity_EH}
\begin{eqnarray}
\left. \frac{(E_{j+} )_y +(E_{j-} )_{y} }{\sqrt{\kappa _j } } 
\right| _{x=\xi _{jj'} }
&=&\left. \frac{(E_{j' +} )_y +(E_{j' -} )_y }{\sqrt{\kappa _{j'} } } 
\right| _{x=\xi _{jj'} } , \nonumber \\
&& \\
\left. \frac{(H_{j+} )_z +(H_{j-} )_z }{\sqrt{\kappa _j } } 
\right| _{x=\xi _{jj'} } 
&=&\left. \frac{(H_{j' +} )_z +(H_{j' -} )_z }{\sqrt{\kappa _j' } } 
\right| _{x=\xi _{jj'} } , \nonumber \\
\end{eqnarray}
\end{subequations}
where $(E_{j\pm } )_y $ and $(H_{j\pm } )_z $ are the $y$- and $z$-components of operators 
$\bm{E}_{j\pm } $ and $\bm{H}_{j\pm } $, respectively, and $\xi _{jj'} $ is the position 
of the interface. 
Using electromagnetic fields 
similar to Eqs. (\ref{electric_field}) and (\ref{magnetic_field}) 
and the wave vectors introduced before, eigenvalues $\alpha _{j\pm } ,\alpha _{j' \pm } $ 
are related to the normalized Fresnel 
coefficients as follows~\footnote{This formulation was also referred to in an unpublished 
work by M.~Matsuoka. }: 
\begin{equation}
\left( 
\begin{array}{c}
\alpha _{j' +} \\
\alpha _{j-} 
\end{array}
\right) =\left( 
\begin{array}{cc}
\tilde{t}_{jj'} & -\tilde{r}_{jj'} \\
\tilde{r}_{jj'} & \tilde{t}_{jj'} 
\end{array}
\right) \left( 
\begin{array}{c}
\alpha _{j+} \\
\alpha _{j' -} 
\end{array}
\right) , \label{transformation_matrix}
\end{equation}
where 
\begin{subequations}
\begin{eqnarray}
\tilde{t}_{jj'} &=&\frac{2\sqrt{\kappa _j \kappa _{j'} } }{\kappa _j +\kappa _{j'} } , 
\label{normalized_Fresnel_t} \\
\tilde{r}_{jj'} &=&\frac{\kappa _j -\kappa _{j'} }{\kappa _j +\kappa _{j'} } \label{normalized_Fresnel_r}
\end{eqnarray}
\end{subequations}
and the phase factor $\exp (\pm i\kappa _j \xi _{jj'} )$ 
is renormalized in the operators of the area $j$, that is, 
$\alpha _{j\pm }\rightarrow \alpha _{j\pm } \exp (\mp i\kappa _j \xi _{jj'} )$. 
It is clear from Eqs. (\ref{normalized_Fresnel_t}) and (\ref{normalized_Fresnel_r})
that $\tilde{t}_{jj'} ^2 +\tilde{r}_{jj'} ^2 =1$ holds.

Introducing $|\psi \rangle =|\alpha _{j+} \rangle _{j+} |\alpha _{j-} \rangle _{j-} 
|\alpha _{j' +} \rangle _{j' +} |\alpha _{j' -} \rangle _{j' -} $, which satisfies 
$\hat{a}_{i\pm } |\psi \rangle =\alpha _{i\pm } |\psi \rangle $ for $i=j,j' $, 
we rewrite Eq. (\ref{transformation_matrix}) as follows: 
\begin{eqnarray}
\langle \psi |\left( 
\begin{array}{c}
\hat{a}_{j' +} \\
\hat{a}_{j-} 
\end{array}
\right) |\psi \rangle &=&\langle \psi |\left( 
\begin{array}{cc}
\tilde{t}_{jj'} & -\tilde{r}_{jj'} \\
\tilde{r}_{jj'} & \tilde{t}_{jj'} 
\end{array}
\right) \left( 
\begin{array}{c}
\hat{a}_{j+} \\
\hat{a}_{j' -} 
\end{array}
\right) |\psi \rangle \nonumber \\
&=&\langle \psi |\hat{\mathcal{U}}_{jj'} \left( 
\begin{array}{c}
\hat{a}_{j+} \\
\hat{a}_{j' -} 
\end{array}
\right) \hat{\mathcal{U}}_{jj'} ^\dagger |\psi \rangle , \label{Fresnel_operator}
\end{eqnarray}
where 
\begin{equation}
\hat{\mathcal{U}} _{jj'} 
=\exp [\vartheta _{jj'} (\hat{a}_{j+} ^\dagger \hat{a}_{j' -} 
-\hat{a}_{j+} \hat{a}_{j' -} ^\dagger )], \label{BS_operator}
\end{equation}
is the beam splitter operator (see Appendix \ref{BS_op}). 
Here, $\vartheta _{jj'} $ is a real parameter and is related 
to the normalized Fresnel coefficients as $\tan \vartheta _{jj'} 
=\tilde{r}_{jj'} /\tilde{t}_{jj'} $. 
The parameter $\vartheta _{jj'} $ should be distinguished 
from the incident and refraction angles, $\theta _j $ and $\theta _{j'} $. 
Equation (\ref{Fresnel_operator}) shows the relation between the operators 
of the optical fields. 
\begin{subequations}
\begin{eqnarray}
\hat{a}_{j' +} &=&\hat{\mathcal{U}}_{jj'} \hat{a}_{j+} \hat{\mathcal{U}}_{jj'} ^\dagger , \\
\hat{a}_{j-} &=&\hat{\mathcal{U}}_{jj'} \hat{a}_{j' -} \hat{\mathcal{U}}_{jj'} ^\dagger . 
\end{eqnarray}
\end{subequations}

The state after interference is obtained by transforming Eq. (\ref{BS_operator}). 
\begin{eqnarray}\
\lefteqn{|\alpha _{j+} \rangle _{j+} |\alpha _{j' -} \rangle _{j' -} 
\rightarrow \hat{\mathcal{U}}_{jj'} |\alpha _{j+} \rangle _{j' +} |\alpha _{j' -} \rangle _{j-} } 
\nonumber \\
&=&|\tilde{t}_{jj'} \alpha _{j+} +\tilde{r}_{jj'} \alpha _{j' -} \rangle _{j' +} 
|-\tilde{r}_{jj'} \alpha _{j+} +\tilde{t}_{jj'} \alpha _{j' -} \rangle _{j-} , \nonumber \\
\end{eqnarray}
where $\tilde{t}_{jj' } =\cos \vartheta _{jj' } $ and $\tilde{r}_{jj' } =\sin \vartheta _{jj'} $ 
are considered. 
It is clear that we obtain $\tilde{t}_{jj'} =1 $ and $\tilde{r}_{jj'} =0 $ when $n_{j'} =n_j $. 
This special case means that the optical field travels in a straight line 
in a homogeneous dielectric, without refraction.

The normalized Fresnel coefficients are related to the conventional Fresnel coefficients 
in a simple way, 
\begin{subequations}
\begin{eqnarray}
\tilde{t}_{jj'} &=&\sqrt{\frac{\kappa _{j'} }{\kappa _j } } t_{jj'} 
=\sqrt{\frac{\kappa _j }{\kappa _{j'} } } t' _{jj'} , \\
\tilde{r}_{jj'} &=&r_{jj'} =-r' _{jj'} 
\end{eqnarray}
\end{subequations}
for the coefficients $t_{jj'} $ ($t' _{jj'} $) related to the transmission 
$j+\rightarrow j' +$ ($j' -\rightarrow j-$) and $r_{jj'} $ ($r' _{jj'} $) related 
to the reflection $j+\rightarrow j-$ ($j' -\rightarrow j' +$).

\subsection{Total reflection case}

Another special case is when the optical field $j$ comes in 
at the critical angle 
\begin{equation}
\theta _c =\sin ^{-1} (n_{j'} /n_j ) \label{critical_angle}
\end{equation}
for $n_j >n_{j'} $. 
This is the well-known total reflection case, and the refraction angle 
is $\theta _{j'} =\pi /2$. 
This means that the lateral component of the wave vector in area $j'$ 
vanishes as $\kappa _{j'} =0$. In this case, we can easily calculate 
$\tilde{t}_{jj'} =0$ and $\tilde{r}_{jj'} =1$ 
from Eqs. (\ref{normalized_Fresnel_t}) and (\ref{normalized_Fresnel_r}). 
When $\theta _c <\theta <\pi /2$, the refraction angle is mathematically shown 
as $\theta _{j'} =\pi /2 -i\varphi _{j' } $ with a real number $\varphi _{j' } $ 
\cite{Navasquillo89}. We then have 
\begin{equation}
\kappa _{j'} =n_{j'} k_0 \cos \theta _{j'} =in_{j'} k_0 \sinh \varphi _{j' } 
\equiv i\gamma _{j'} . \label{kappa_2}
\end{equation}
The real and imaginary part of the wave vector is related to the wave number and effective loss, 
respectively \cite{Lodenquai91}, 
and Eq. (\ref{kappa_2}) shows that the wave number along the $x$-direction is zero 
in area $j'$. This means that we can also regard $\kappa _{j'} $ as zero in Eqs. 
(\ref{normalized_Fresnel_t}) and (\ref{normalized_Fresnel_r}) and again we find 
the total reflection case.

So far, according to the present model, no optical field exists in the area $j'$ 
because it is totally reflected at the interface. 
In practice, however, a certain amount of the optical field, 
as far as the wavelength distance, can leak in to area $j'$. 
This leakage is referred to 
as the evanescent field \cite{Snyder_text} and we can clarify it using the uncertainty relation. 
As stated above, for the total reflection case, the incident angle satisfies 
the inequality $\theta _c \leq \theta _j <\pi /2$. The inequality is reduced to 
\begin{equation}
0<\kappa _j \leq \kappa _c =n_j k_0 \cos \theta _c \label{inequality_kappa1}
\end{equation}
with an $x$-component of the wave vector in area $j$. 
This inequality (\ref{inequality_kappa1}) shows that the totally reflected optical field 
displays a fluctuation $\Delta \kappa _j =\kappa _c $. 
Using de Broglie's relation, 
\begin{equation}
\Delta p_x =\hbar \Delta \kappa _j =k_0 \sqrt{n_j ^2 -n_{j'} ^2 } , 
\end{equation}
where Eq. (\ref{critical_angle}) is also considered. 
From the uncertainty relation, 
\begin{equation}
\Delta q_x \simeq \frac{\hbar }{\Delta p_x } =\frac{\lambda _0 }{2\pi n_j \sqrt{2\Delta _r } } , 
\end{equation}
where $\lambda _0 =2\pi /k_0 $ is the wave length of the optical field 
in vacuum and 
\begin{equation}
\Delta _r =\frac{n_j ^2 -n_{j'} ^2 }{2n_j ^2 } 
\end{equation}
is the relative index difference. Introducing $n_j =1.45$ and $\Delta _r =0.01$  
as typical parameters, we obtain $\Delta q_x \simeq \lambda _0 /1.29$. 
This result shows that the optical field can proceed into the area prohibited 
by the total reflection scheme, by the degree of a wavelength.

\section{Propagation of the optical field in slab waveguides}
\label{waveguide}
In the previous sections, we studied the propagation of an optical field 
in a homogeneous dielectric and its behavior at an interface. 
By combining these procedures, we can describe the behavior of optical fields 
in an inhomogeneous dielectric, such as waveguides.

\subsection{Step-index slab waveguide}

The simplest example is the step-index (SI) slab waveguide. 
It consists of a homogeneous core and cladding, 
with refractive indices of $n_1 $ and $n_2 (<n_1)$, respectively. 
The waveguide propagates the optical field along the longitudinal direction 
and confines it within the cross section. 
This mechanism is achieved by total (internal) reflection at the core-cladding interface.

\begin{figure}[t]
\centering 
\includegraphics[bb=15 165 525 480, width=\linewidth, clip]{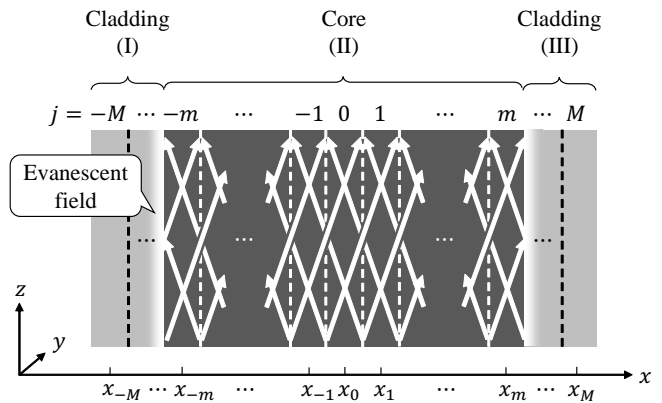}
\caption{\label{step-index} Step-index slab waveguide in a symmetric structure. 
The core and cladding are divided into layers with a common thickness $\Delta x$ 
representing the spatial dependence of the optical field. The optical field 
is chiefly confined in the core, but part of it leaks to the cladding 
as an evanescent field. }
\end{figure}

A schematic of the SI slab waveguide is shown 
in Fig. \ref{step-index}. Here, a symmetrical structure, where 
refractive indices of left- and right-side cladding are identical, is considered. 
To represent the spatial dependence of the optical field in the waveguide, 
the core and cladding are divided into stacked layers, the thicknesses of which is 
denoted by $\Delta x$. The order of layers is distinguished 
by an index $j$, and the $j$th layer is located in the range 
\begin{equation}
\left( j-\frac{1}{2} \right) \Delta x\leq x\leq \left( j+\frac{1}{2} \right) \Delta x. 
\label{jth_rectangle}
\end{equation}
We assume that the core and cladding correspond to areas $|j|\leq m$ 
and $m+1\leq |j|\leq M$, respectively. We introduce the center position 
of the $j$th layer as 
\begin{equation}
x_j =j\Delta x. \label{x_j}
\end{equation}
Let the position of the core-cladding interface be $x=x_{\pm m} \pm \Delta x/2=\pm d/2$, 
where double-sign in same order is assumed. 
For convenience we refer to the left cladding, core, and right cladding as 
areas (I), (II), and (III), respectively.

The optical field in the cladding is distributed as an evanescent field. 
The $x$-component of wave vector of the evanescent field is given 
by an imaginary number as shown in Eq. (\ref{kappa_2}). 
This means that the wave vector can be complex. 
Since the operator corresponding to the electromagnetic field 
should be Hermitian, we introduce the vector potential 
in the SI slab waveguide as 
\begin{eqnarray}
\hat{\bm{A}}_j &=&\sqrt{\frac{\hbar }{2\omega \varepsilon _0 V_j } } 
\Big\{ \hat{a}_j ^\dagger \exp [i(\omega t-\bm{k}_j ^\ast \cdot \bm{r})] \nonumber \\
&&\hspace{20mm} +\hat{a}_j \exp [-i(\omega t-\bm{k}_j \cdot \bm{r})] \Big\} \bm{e}_y , 
\label{SI_vector_potential} 
\end{eqnarray}
with $V_j =\Delta x\cdot 1\cdot 2\pi n_z /\beta $. 
The asterisk in Eq. (\ref{SI_vector_potential}) shows the complex conjugate. 
This vector potential gives the electromagnetic fields of a TE mode. 
In areas (I), (II), and (III) wave vectors are given by 
\begin{subequations}
\begin{eqnarray}
\bm{k}^{\rm (I)} _j &=&-i\gamma \bm{e}_x +\beta \bm{e}_z , \\
\bm{k}^{\rm (II)} _{j\pm } &=&\pm \kappa \bm{e}_x +\beta \bm{e}_z , \label{k_II} \\
\bm{k}^{\rm (III)} _j &=&i\gamma \bm{e}_x +\beta \bm{e}_z , 
\end{eqnarray}
\end{subequations}
respectively. It is clear that the absolute values of the $x$-components of $\bm{k}^{\rm (I)} _j $ 
and $\bm{k}^{\rm (III)} _j $ are identical because the structure of the waveguide is symmetrical 
and the incident angles at the interfaces between the core and cladding are identical. 
Optical fields with $x$-components of $\pm \kappa $ are shown by a double-sign notation 
in $\bm{k}^{\rm (II)} _{j\pm } $. 
These fields are spatially superposed resulting in a sinusoidal distribution 
in the core. Note that, at the interface between adjacent layers, optical fields travel 
with no interaction because the core is assumed to be  homogeneous 
and the normalized Fresnel coefficient for the transmission is unity.

The vector potentials in areas (I), (II), and (III) are defined as follows: 
\begin{subequations}
\begin{eqnarray}
\hat{\bm{A}}_j ^{\rm (I)} &=&\sqrt{\frac{\hbar }{2\omega \varepsilon _0 V_j }} 
(U_{tz} ^\ast \hat{a}_j ^{{\rm (I)}\dagger }
+U_{tz} \hat{a}_j ^{\rm (I)} )\exp (\gamma x)\bm{e}_y \nonumber \\
&&\hspace{25mm} (-M\leq j\leq -m-1), \\
\hat{\bm{A}}_j ^{\rm (II)} &=&\frac{\hat{\bm{A}}_{j+} ^{\rm (II)} 
+\hat{\bm{A}}_{j-} ^{\rm (II)} }{\sqrt{2} } \quad (-m\leq j\leq m), \\
\hat{\bm{A}}_j ^{\rm (III)} &=&\sqrt{\frac{\hbar }{2\omega \varepsilon _0 V_j }} 
(U_{tz} ^\ast \hat{a}_j ^{{\rm (III)} \dagger } 
+U_{tz} \hat{a}_j ^{\rm (III)} )\exp (-\gamma x)\bm{e}_y \nonumber \\
&&\hspace{25mm} (m+1\leq j\leq M), 
\end{eqnarray}
\end{subequations}
where $U_{tz} =\exp [-i(\omega t-\beta z)]$ is a temporal- and $z$-dependence 
similar to Eq. (\ref{U_tz}) and 
\begin{eqnarray}
\hat{\bm{A}}_{j\pm } ^{(\rm II)} &=&\sqrt{\frac{\hbar }{2\omega \varepsilon _0 V_j }} 
[U_{tz} ^\ast \hat{a}_{j\pm } ^{{\rm (II)}\dagger } \exp (\mp i\kappa x) \nonumber \\
&&\hspace{20mm} +U_{tz} \hat{a}_{j\pm } ^{\rm (II)} \exp (\pm i\kappa x)]\bm{e}_y 
\end{eqnarray}
with double-sign in same order. Here, $\hat{a}_{j\pm } $ is the operator 
for the optical field with the wave vector (\ref{k_II}). 
Similar to Eqs. (\ref{electric_field}) and (\ref{magnetic_field}),   
the electromagnetic fields in each area are calculated thus: 
\begin{subequations}
\begin{eqnarray}
\hat{\bm{E}}_j ^{\rm (I)} &=&-i\sqrt{\frac{\hbar \omega }{2\varepsilon _0 V_j }} 
(U_{tz} ^\ast \hat{a}_j ^{{\rm (I)}\dagger } 
-U_{tz} \hat{a}_j ^{\rm (I)} )\exp (\gamma x)\bm{e}_y , \\
\hat{\bm{E}}_j ^{\rm (II)} &=&\frac{\hat{\bm{E}}_{j+} ^{\rm (II)} 
+\hat{\bm{E}}_{j-} ^{\rm (II)} }{\sqrt{2} } , \\
\hat{\bm{E}}_j ^{\rm (III)} &=&-i\sqrt{\frac{\hbar \omega }{2\varepsilon _0 V_j }} 
(U_{tz} ^\ast \hat{a}_j ^{{\rm (III)}\dagger } 
-U_{tz} \hat{a}_j ^{\rm (III)} )\exp (-\gamma x)\bm{e}_y , \nonumber \\
&& \\
\hat{\bm{H}}_j ^{\rm (I)} &=&-\frac{i}{\mu _0 } 
\sqrt{\frac{\hbar }{2\omega \varepsilon _0 V_j } } 
[-\beta (U_{tz} ^\ast \hat{a}_j ^{{\rm (I)}\dagger } 
-U_{tz} \hat{a}_j ^{\rm (I)} )\bm{e}_x \nonumber \\
&&\hspace{10mm} +i\gamma (U_{tz} ^\ast \hat{a}_j ^{{\rm (I)}\dagger } 
+U_{tz} \hat{a}_j ^{\rm (I)} )\bm{e}_z ]\exp (\gamma x), \\
\hat{\bm{H}}_j ^{\rm (II)} &=&\frac{\hat{\bm{H}}_{j+} ^{\rm (II)} 
+\hat{\bm{H}}_{j-} ^{\rm (II)} }{\sqrt{2} } , \\
\hat{\bm{H}}_j ^{\rm (III)} &=&-\frac{i}{\mu _0 } \sqrt{\frac{\hbar }{2\omega \varepsilon _0 V_j } } 
[-\beta (U_{tz} ^\ast \hat{a}_j ^{{\rm (III)} \dagger }
-U_{tz} \hat{a}_j ^{\rm (III)} )\bm{e}_x \nonumber \\
&&\hspace{10mm} -i\gamma (U_{tz} ^\ast \hat{a}_j ^{{\rm (III)}\dagger }
+U_{tz} \hat{a}_j ^{\rm (III)} )\bm{e}_z ]\exp (-\gamma x), \nonumber \\
\end{eqnarray}
\end{subequations}
where 
\begin{subequations}
\begin{eqnarray}
\hat{\bm{E}}_{j\pm } ^{\rm (II)} &=&-i\sqrt{\frac{\hbar \omega }
{2\varepsilon _0 V_j }} 
[U_{tz} ^\ast \hat{a}_{j\pm } ^{{\rm (II)}\dagger } \exp (\mp i\kappa x) \nonumber \\
&&\hspace{5mm} -U_{tz} \hat{a}_{j\pm } ^{\rm (II)} \exp (\pm i\kappa x)]\bm{e}_y , 
\label{E^(II)} \\
\hat{\bm{H}}_{j\pm } ^{\rm (II)} &=&-\frac{i}{\mu _0 } 
\sqrt{\frac{\hbar }{2\omega \varepsilon _0 V_j }} 
[U_{tz} ^\ast \hat{a}_{j\pm } ^{{\rm (II)}\dagger } \exp (\mp i\kappa x) \nonumber \\
&&\hspace{5mm} -U_{tz} \hat{a}_{j\pm } ^{\rm (II)} \exp (\pm i\kappa x)] 
(-\beta \bm{e}_x \pm \kappa \bm{e}_z ) \label{H^(II)}
\end{eqnarray}
\end{subequations}
with double-sign in same order.

Here we introduce a spatially-averaged operator in the $j$th layer. 
For areas (I), (II), and (III), we have 
\begin{subequations}
\begin{eqnarray}
c^{\rm (I)} _j &=&\frac{1}{\Delta x} \int _{(j-\frac{1}{2})\Delta x } ^{(j+\frac{1}{2})\Delta x} 
\hat{a}^{\rm (I)} _j \exp (\gamma x)dx \nonumber \\
&\simeq &\hat{a}_j ^{\rm (I)} \exp (\gamma x_j ), \\
\hat{c}_j ^{\rm (II)} &=&\frac{1}{\Delta x} \int _{(j-\frac{1}{2} )\Delta x} 
^{(j+\frac{1}{2} )\Delta x} \frac{\hat{a}_{j+} ^{\rm (II)} \exp (i\kappa x) 
+\hat{a}_{j-} ^{\rm (II)} \exp (-i\kappa x) }{\sqrt{2}} dx 
\nonumber \\
&\simeq &\frac{\hat{a}_{j+} ^{\rm (II)} \exp (i\kappa x_j )
+\hat{a}_{j-} ^{\rm (II)} \exp (-i\kappa x_j )}
{\sqrt{2}} , \\
c^{\rm (III)} _j &=&\frac{1}{\Delta x} \int _{(j-\frac{1}{2})\Delta x } ^{(j+\frac{1}{2})\Delta x} 
\hat{a}^{\rm (III)} _j \exp (-\gamma x)dx \nonumber \\
&\simeq &\hat{a}_j ^{\rm (III)} \exp (-\gamma x_j ) 
\end{eqnarray}
\end{subequations}
with a small $\Delta x$. 
In area (II) in particular, this spatially-averaged operator characterizes 
a standing wave formed by the interference between the electric (magnetic) fields 
$\hat{\bm{E}}_{j+} ^{\rm (II)} $ and $\hat{\bm{E}}_{j-} ^{\rm (II)} $ 
($\hat{\bm{H}}_{j+} ^{\rm (II)} $ and $\hat{\bm{H}}_{j+} ^{\rm (II)} $). 
It is obvious that the spatially-averaged operator 
$\hat{c}_j ^{\rm (K)} (\textrm{K}=\textrm{I},\textrm{II},\textrm{III})$ satisfies 
a commutation relation: 
\begin{subequations}
\begin{eqnarray}
&&[\hat{c}^{\rm (I)} _j ,\hat{c}^{{\rm (I)} \dagger } _j ]=\exp (2\gamma x_j ), 
\label{commutation_I} \\
&&[\hat{c}_j ^{\rm (II)} ,\hat{c}_j ^{{\rm (II)}\dagger } ]=1, \label{commutation_II}\\
&&[\hat{c}^{\rm (III)} _j ,\hat{c}^{{\rm (III)} \dagger } _j ]=\exp (-2\gamma x_j ). 
\label{commutation_III}
\end{eqnarray}
\end{subequations}
The right-hand side of Eqs. (\ref{commutation_I}) and (\ref{commutation_III}) are 
not unity because the optical fields 
in areas (I) and (III) are the evanescent fields, respectively.

We can evaluate the number of photons in the $j$th layer with spatially-averaged operators: 
$\hat{N}^{\rm (K)} _j =\hat{c}^{{\rm (K)} \dagger } _j \hat{c}^{\rm (K)} _j 
(\textrm{K}=\textrm{I}, \textrm{II}, \textrm{III})$. 
The momentum of the field in the area (K) is calculated 
by the $z$-component of the Poynting vector for a small $\Delta x$ \cite{Louisell_text}. 
\begin{eqnarray}
\left. \hat{G}^{\rm (K)} _j \right| _z &=&\frac{1}{c^2 } \int (\hat{\bm{E}}^{\rm (K)} _j 
\times \hat{\bm{H}}^{\rm (K)} _j )\cdot \bm{e}_z dV \nonumber \\
&\simeq &\frac{\hbar \beta }{2} (\hat{c}^{\rm (K)} _j \hat{c}^{{\rm (K)} \dagger } _j 
+\hat{c}^{{\rm (K)} \dagger } _j \hat{c}^{\rm (K)} _j ), 
\label{energy_flow_j}
\end{eqnarray}
where the above integration is calculated over the unit volume of the $j$th layer $V_j $ 
for a small $\Delta x$. 
Using the commutation relation, it is clear that energy flow in the $j$th layer is 
proportional to the number of photons in the $j$th layer.  
Total momentum in the waveguide is obtained by 
\begin{equation}
\hat{G}_z =\sum _{j=-M} ^M  \left. \hat{G}^{\rm (K)} _j \right| _z . \label{energy_flow}
\end{equation}
This operator is related to the energy flow along the $z$-direction 
and is discussed later.

Let us consider the expectation values of electromagnetic fields. In the present scheme, 
we assume that the field is in a coherent state over the complete cross-section 
of the optical waveguide. This assumption is valid because the distribution 
of the electromagnetic field is static in cross-section in SI optical waveguides. 
The total state of the optical field is described as 
\begin{equation}
|\alpha \rangle =\bigotimes _{j=-M} ^M |\alpha ^{\rm (K)} _j \rangle _j , 
\end{equation}
where $\textrm{K}=\textrm{I}$ for $-M\leq j\leq -m-1$, $\textrm{K}=\textrm{II}$ 
for $-m\leq j\leq m$, and $\textrm{K}=\textrm{III}$ for $m+1\leq j\leq M$. 
Note that, in area (II), we should distinguish optical fields 
with wave vectors $\bm{k}^{\rm (II)} _{j\pm } $ as different spatial modes, 
then we have $|\alpha ^{\rm (II)} _j \rangle _j =|\alpha ^{\rm (II)} _{j+} \rangle _{j+} 
\otimes |\alpha ^{\rm (II)} _{j-} \rangle _{j-} $.

Also, we introduce eigenvalues for coherent state optical field operator. For each area, 
\begin{subequations}
\begin{eqnarray}
\hat{a}_j ^{\rm (K)} |\alpha _j ^{\rm (K)} \rangle _j 
&=&\alpha _j ^{\rm (K)} |\alpha _j ^{\rm (K)} \rangle _j 
\quad (\textrm{K}=\textrm{I},\textrm{III}), \label{eigenvalue_I-III} \\
\hat{a}_{j\pm } ^{\rm (II)} |\alpha _{j\pm } ^{\rm (II)} \rangle _{j\pm } 
&=&\alpha _{j\pm } ^{\rm (II)} |\alpha _{j\pm } ^{\rm (II)} \rangle _{j\pm } \label{eigenvalue_II}
\end{eqnarray}
\end{subequations}
with double-sign in same order. We can obtain expectation values 
of electromagnetic fields in the $j$th layer as follows for $\textrm{K}=\textrm{I}, 
\textrm{II}, \textrm{III}$: 
\begin{subequations}
\begin{eqnarray}
\bm{E}_j ^{\rm (K)} &=& _j \langle \alpha ^{\rm (K)} _j |\hat{\bm{E}}_j ^{\rm (K)} 
|\alpha ^{\rm (K)} _j \rangle _j , \\
\bm{H}_j ^{\rm (K)} &=& _j \langle \alpha ^{\rm (K)} _j |\hat{\bm{H}}_j ^{\rm (K)} 
|\alpha ^{\rm (K)} _j \rangle _j . 
\end{eqnarray}
\end{subequations}

From the continuity of the electromagnetic fields, 
we can find the relation between the eigenvalues of adjacent layers in each area. 
At the interface between $(j-1)$th and $j$th layers in area (II), 
for example, we have $\bm{E}_{j-1} ^{\rm (II)} (x_{j-1} +\Delta x/2)
=\bm{E}_j ^{\rm (II)} (x_{j-1} +\Delta x/2)$ for $-m+1\leq j\leq m$. 
This results in $\alpha _{(j-1)\pm } ^{\rm (II)} =\alpha _{j\pm } ^{\rm (II)} $. 
It shows that eigenvalues are the same over all of area (II).  In other words, 
they are independent of $j$. 
Let $\alpha _{j+} ^{\rm (II)} $ and $\alpha _{j-} ^{\rm (II)} $ 
be $\alpha _+ ^{\rm (II)} $ and $\alpha _- ^{\rm (II)} $. 
Similarly, eigenvalues in areas (I) and (III) are also independent of $j$, 
and we introduce them as $\alpha ^{\rm (I)} $ and $\alpha ^{\rm (III)} $, respectively.

Eigenvalues in area (II) correspond to the superposition weighting coefficients 
found in classical waveguide optics. 
We obtain TE even and odd modes for cases of $\alpha ^{\rm (II)} _+=\alpha _- ^{\rm (II)} 
=\alpha ^{\rm (II)} _{\rm even} $ 
and $\alpha  ^{\rm (II)} _+ =-\alpha ^{\rm (II)} _- 
=\alpha ^{\rm (II)} _{\rm odd} $, respectively. 
Using these symbols, we can write eigenvalues of $\hat{c}^{\rm (K)} _j $ 
for $\textrm{K}=\textrm{I, II, III}$ as follows: 
\begin{subequations}
\begin{eqnarray}
\hat{c}^{\rm (I)} _j |\alpha ^{\rm (I)} _j \rangle _j 
&=&\exp (\gamma x_j )\alpha ^{\rm (I)} |\alpha _j ^{\rm (I)} \rangle _j , \\
\hat{c}^{\rm (II)} _j |\alpha ^{\rm (II)} _j \rangle _j 
&=&\left\{ 
\begin{array}{l}
\sqrt{2}\alpha ^{\rm (II)} _{\rm even} \cos \kappa x_j |\alpha ^{\rm (II)} _j \rangle _j \\
\hspace{15mm} \textrm{(TE even mode)} \\
\sqrt{2}i\alpha ^{\rm (II)} _{\rm odd} \sin \kappa x_j |\alpha ^{\rm (II)} _j \rangle _j \\
\hspace{15mm} \textrm{(TE odd mode)} 
\end{array}
\right. , \\
\hat{c}^{\rm (III)} _j |\alpha ^{\rm (III)} _j \rangle _j 
&=&\exp (-\gamma x_j )\alpha ^{\rm (III)} |\alpha ^{\rm (III)} _j \rangle _j . 
\end{eqnarray}
\end{subequations}

With the above relations, we have 
\begin{subequations}
\begin{eqnarray}
_j \langle \alpha ^{\rm (I)} _j |\hat{c}^{{\rm (I)} \dagger } _j 
\hat{c}^{\rm (I)} _j |\alpha ^{\rm (I)} _j \rangle _j &=&\left| \alpha ^{\rm (I)} \right| ^2 
\exp (2\gamma x_j ), \\
_j \langle \alpha ^{\rm (II)} _j |\hat{c}^{{\rm (II)} \dagger } _j 
\hat{c}^{\rm (II)} _j |\alpha ^{\rm (II)} _j \rangle _j 
&=&\left\{ 
\begin{array}{l}
2\left| \alpha ^{\rm (II)} _{\rm even} \right| ^2 \cos ^2 \kappa x_j \\
\hspace{5mm} (\textrm{TE even mode)} \\
2\left| \alpha ^{\rm (II)} _{\rm odd} \right| ^2 \sin ^2 \kappa x_j \\
\hspace{5mm} (\textrm{TE odd mode)} 
\end{array}
\right. , \nonumber \\
&& \\
\langle \alpha ^{\rm (III)} _j |\hat{c}^{{\rm (III)} \dagger } _j 
\hat{c}^{\rm (II)} _j |\alpha ^{\rm (III)} _j \rangle _j &=&\left| \alpha ^{\rm (III)} \right| ^2 
\exp (-2\gamma x_j ),
\end{eqnarray}
\end{subequations}
which gives the mean number of photons in the $j$th layer. These values are related to 
the probability of existence of a photon in the $j$th layer 
when the amplitude of the coherent state is small such as $|\alpha |<1$.

Eigenvalues $\alpha ^{\rm (I)} $, $\alpha ^{\rm (II)} _{\rm even} $ 
($\alpha ^{\rm (II)} _{\rm odd} $), and $\alpha ^{\rm (III)} $ 
are associated with each other through continuity condition 
of the tangential components of the electric or magnetic fields at the interfaces 
between areas (I)-(II) ($x=-d/2$) and (II)-(III) ($x=d/2$), respectively. 
From these conditions, similar to classical waveguide optics, we find that the 
eigenvalues are related (see Appendix \ref{Appendix_C}): 
\begin{subequations}
\label{ratio_coefficients}
\begin{eqnarray}
\frac{\alpha ^{\rm (I)} }{\alpha ^{\rm (II)} _{\rm even} } 
&=&\frac{\alpha ^{\rm (III)} }{\alpha ^{\rm (II)} _{\rm even} } 
=\sqrt{2} \exp \left( \frac{\gamma d}{2} \right) \cos \frac{\kappa d}{2} \nonumber \\
&&\hspace{25mm} (\textrm{TE even mode}), \label{ratio_coefficients_even} \\
-\frac{\alpha ^{\rm (I)} }{\alpha ^{\rm (II)} _{\rm odd} } 
&=&\frac{\alpha ^{\rm (III)} }{\alpha ^{\rm (II)} _{\rm odd} } 
=\sqrt{2} i\exp \left( \frac{\gamma d}{2} \right) \sin \frac{\kappa d}{2} \nonumber \\
&&\hspace{25mm} (\textrm{TE odd mode}). \label{ratio_coefficients_odd} 
\end{eqnarray}
\end{subequations}
The eigenvalue $\alpha ^{\rm (II)} _{\rm even} $ or $\alpha ^{\rm (II)} _{\rm odd} $ 
should be determined through normalization. 
This condition is related to the amount of total energy in the waveguide. 
Energy flow of the optical field is obtained from the expectation value 
of Eq. (\ref{energy_flow}). 
\begin{equation}
S=c^2 \langle \alpha |\hat{G}_z |\alpha \rangle . 
\end{equation}
Then $\alpha ^{\rm (I)} $ and $\alpha ^{\rm (III)} $ can be determined 
with Eqs. (\ref{ratio_coefficients_even}) and (\ref{ratio_coefficients_odd}), respectively.

Using eigenvalues, the spatially-averaged electric field in the $j$th layer is calculated. 
\begin{equation}
\bm{\mathcal{E}}_j ^{\rm (K)} =\frac{1}{\Delta x} 
\int _{(j-\frac{1}{2})\Delta x} ^{(j+\frac{1}{2})\Delta x} \bm{E}^{\rm (K)} _j dx 
\quad (\textrm{K}=\textrm{I, II, or III}). 
\end{equation}
Distinct representations are obtained for a small $\Delta x$ as follows: 
\begin{subequations}
\begin{eqnarray}
\bm{\mathcal{E}}_j ^{\rm (I)} &\simeq &-i\sqrt{\frac{\hbar \omega }{2\varepsilon _0 V_j }} 
(U_{tz} ^\ast \alpha ^{{\rm (I)} \ast } -U_{tz} \alpha ^{\rm (I)} )
\exp (\gamma x_j )\bm{e}_y , \\
\bm{\mathcal{E}}_j ^{\rm (II)} &\simeq &\left\{ 
\begin{array}{l}
-i\sqrt{\displaystyle \frac{\hbar \omega }{\varepsilon _0 V_j } } 
(U_{tz} ^\ast \alpha _{\rm even} ^{{\rm (II)} \ast } 
-U_{tz} \alpha _{\rm even} ^{\rm (II)} )\cos \kappa x_j \bm{e}_y \\
\hspace{35mm} \textrm{(TE even mode)} \\
-\sqrt{\displaystyle \frac{\hbar \omega }{\varepsilon _0 V_j } } 
(U_{tz} ^\ast \alpha _{\rm odd} ^{{\rm (II)} \ast } 
+U_{tz} \alpha _{\rm odd} ^{\rm (II)} )\sin \kappa x_j \bm{e}_y \\
\hspace{35mm} \textrm{(TE odd mode)} 
\end{array}
\right. , \nonumber \\
&& \\
\bm{\mathcal{E}}_j ^{\rm (III)} &\simeq &-i\sqrt{\frac{\hbar \omega }{2\varepsilon _0 V_j }} 
(U_{tz} ^\ast \alpha ^{{\rm (III)} \ast } -U_{tz} \alpha ^{\rm (III)} )
\exp (-\gamma x_j )\bm{e}_y . \nonumber \\
\end{eqnarray}
\end{subequations}
In area (II), the electric fields $\hat{\bm{E}}^{\rm (II)} _{j+} $ 
and $\hat{\bm{E}}^{\rm (II)} _{j-} $ interfere with each other and we find 
that sinusoidal distributions appear for both TE even and odd modes as a result.

The spatially-averaged magnetic field is similarly calculated. 
\begin{equation}
\bm{\mathcal{H}}_j ^{\rm (K)} =\frac{1}{\Delta x} 
\int _{(j-\frac{1}{2})\Delta x} ^{(j+\frac{1}{2})\Delta x} \bm{H}^{\rm (K)} _j dx 
\quad (\textrm{K}=\textrm{I, II, or III}), 
\end{equation}
and 
\begin{subequations}
\begin{eqnarray}
\bm{\mathcal{H}}_j ^{\rm (I)} &\simeq &-\frac{i}{\mu _0 } 
\sqrt{\frac{\hbar }{2\omega \varepsilon _0 V_j } } 
[-\beta (U_{tz} ^\ast \alpha ^{{\rm (I)} \ast } -U_{tz} \alpha ^{\rm (I)} )
\bm{e}_x \nonumber \\
&&\hspace{5mm} +i\gamma (U_{tz} ^\ast \alpha ^{{\rm (I)} \ast } 
+U_{tz} \alpha ^{\rm (I)} )\bm{e}_z ]\exp (\gamma x_j ), \\
\bm{\mathcal{H}}_j ^{\rm (II)} &\simeq &\left\{ 
\begin{array}{l}
-\displaystyle \frac{i}{\mu _0 } \sqrt{\displaystyle \frac{\hbar }{\omega \varepsilon _0 V_j } } \\
\times [-\beta (U_{tz} ^\ast \alpha _{\rm even} ^{{\rm (II)} \ast } 
-U_{tz} \alpha _{\rm even} ^{\rm (II)} ) \cos \kappa x_j \bm{e}_x \\
\hspace{5mm} -i\kappa (U_{tz} ^\ast \alpha ^{{\rm (II)} \ast } _{\rm even} 
+U_{tz} \alpha ^{\rm (II)} _{\rm even} )\sin \kappa x_j \bm{e}_z ] \\
\hspace{35mm} \textrm{(TE even mode)} \\
-\displaystyle \frac{i}{\mu _0 } \sqrt{\displaystyle \frac{\hbar }{\omega \varepsilon _0 V_j } } \\
\times [i\beta (U_{tz} ^\ast \alpha _{\rm odd} ^{{\rm (II)} \ast } 
+U_{tz} \alpha _{\rm odd} ^{\rm (II)} ) \sin \kappa x_j \bm{e}_x \\
\hspace{5mm} +\kappa (U_{tz} ^\ast \alpha ^{{\rm (II)} \ast } _{\rm odd} 
-U_{tz} \alpha ^{\rm (II)} _{\rm odd} )\cos \kappa x_j \bm{e}_z ] \\
\hspace{35mm} \textrm{(TE odd mode)} 
\end{array}
\right. , \nonumber \\
&& \\
\bm{\mathcal{H}}_j ^{\rm (III)} &\simeq &-\frac{i}{\mu _0 } 
\sqrt{\frac{\hbar }{2\omega \varepsilon _0 V_j } } 
[-\beta (U_{tz} ^\ast \alpha ^{{\rm (III)} \ast } 
-U_{tz} \alpha ^{\rm (III)} )\bm{e}_x \nonumber \\
&&\hspace{5mm} -i\gamma (U_{tz} ^\ast \alpha ^{{\rm (III)} \ast } 
+U_{tz} \alpha ^{\rm (III)} )\bm{e}_z ]\exp (-\gamma x_j ). \nonumber \\
\end{eqnarray}
\end{subequations}
The representations obtained here characterize the quantum properties 
of the optical field. For a certain amount of amplitude in the coherent state, however, 
they are consistent with representations in classical waveguide optics.

From the continuity conditions an equation that gives the propagation constant $\beta $ 
is also derived for TE even and odd modes (see Appendix \ref{Appendix_C}): 
\begin{equation}
\gamma =\left\{ 
\begin{array}{ll}
\kappa \tan \displaystyle \frac{\kappa d}{2} & (\textrm{TE even mode}) \\
-\kappa \cot \displaystyle \frac{\kappa d}{2} & (\textrm{TE odd mode})
\end{array}
\right. . \label{eigenvalue_equation}
\end{equation}
Since the propagation constant $\beta $ is mathematically the eigenvalue of a wave equation 
\cite{Snyder_text}, the above equation (\ref{eigenvalue_equation}) is referred to as 
the eigenvalue equation in waveguide optics. Note that the eigenvalue here 
should be distinguished from eigenvalues for the coherent state 
in Eqs. (\ref{eigenvalue_I-III}) and (\ref{eigenvalue_II}). 
Since both $\kappa $ and $\gamma $ are functions of $\beta $, 
Eq. (\ref{eigenvalue_equation}) is a transcendental equation with respect to $\beta $. 
This is also consistent with classical waveguide optics.

\subsection{Graded-index slab waveguide}

\begin{figure}[t]
\centering 
\includegraphics[bb=60 35 695 515, width=\linewidth, clip]{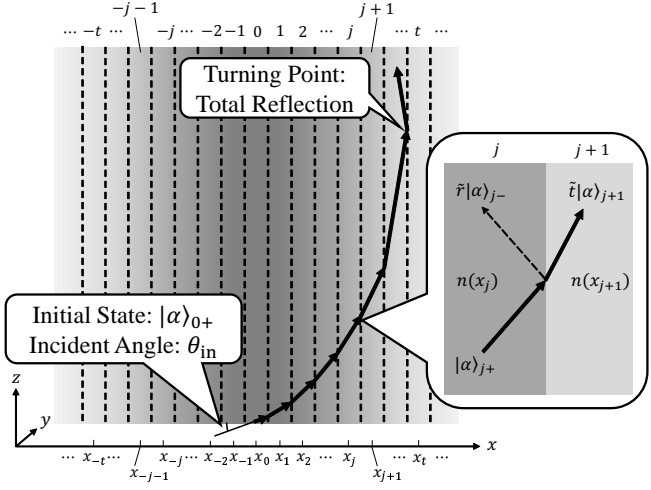}
\caption{\label{graded-index} Graded-index slab waveguide in a symmetric structure. 
A power-law profile of the refractive index is considered. 
The core is divided into layers to represent the spatial dependence of the optical field. 
The optical field is supposed to enter at the $j$th layer with the incident angle $\theta _{in} $ 
(here, $j=0$ is indicated). The optical field travels with a trajectory. }
\end{figure}

Another interesting example is the graded-index (GI) slab waveguide. 
The propagation properties of GI slab waveguides have often been studied 
within geometric optics \cite{Snyder_text}. 
Here we investigate it using the quantum-optical method.

A schematic of the GI slab waveguide is shown in Fig. \ref{graded-index}. 
In contrast to the SI case, the refractive index of the core is dependent 
upon the position in the cross-section. 
We assume that the optical field travels in the core, 
without leakage to the cladding. Let us consider a power-law profile: 
\begin{equation}
n(x)=n_1 \sqrt{1-(g|x|)^q } , \label{power-law profile}
\end{equation}
where $n_1 $ is a refractive index at the center of the core, 
and $g=2\sqrt{2\Delta _r }/d $ is referred to as a focusing constant. 
The width of the core is denoted by $d$. The parameter $q$ characterizes the shape 
of the index distribution. Here, we study the case when $q=2$, 
the parabolic index waveguide that is important for optical communications 
\footnote{The optimum distribution of the refractive index is known 
as $n(x)=n_1 \mathrm{sech} (g|x|)$. It was particularly discussed in the following article: 
S.~Kawakami and J.~Nishizawa, IEEE Trans. Microwave Theory Tech., \textbf{MTT-16}, 
814 (1968). }.

The core is divided into layers with a common thickness $\Delta x$, 
similar to the SI case. In the present quantum scheme 
this multi-layer division enables us to take account not only of variations in 
refractive index, but also of the spatial dependence of the optical field. 
Layers are identified by the index $j$, 
and the center position of the $j$th layer is $x_j =j\Delta x$, similar to the SI case. 
The refractive index of the $j$th layer is represented by $n_j \equiv n(x_j )$. 
The electromagnetic field and the state of the optical fields in the $j$th layer are 
denoted by $\hat{\bm{E}}_{j\pm } ,\hat{\bm{H}}_{j\pm } $ and $|\alpha _{j\pm } \rangle _{j\pm } $, 
respectively. Note here that optical fields with a wave vector consisting 
of positive and negative $x$-components are distinguished 
by the notation $j\pm $. As discussed later, the field is 
transmitted with no attenuation, and we can write 
the amplitude of the coherent state as a value independent of the spatial index $j$, 
that is, $\alpha _{j+} \equiv \alpha $.

Again we consider a TE mode; the state of optical field is 
initially in the $j$th layer at an angle $\theta _{\rm in} $ 
with respect to the $x$-axis. We assume that the $x$-component of the wave vector 
is positive and the initial state is denoted by $|\alpha \rangle _{j+} $. 
Let the wave vector in the $j$th layer be $\bm{k}_{j\pm } 
=\kappa _j \bm{e}_x \pm \beta \bm{e}_z $. 
The corresponding electromagnetic fields are represented by 
\begin{subequations}
\begin{eqnarray}
\hat{\bm{E}}_{j\pm } &=&-i\sqrt{\frac{\hbar \omega }
{2\varepsilon _0 V_j }} 
[U_{tz} ^\ast \hat{a}_{j\pm } ^\dagger \exp (\mp i\kappa _j x) \nonumber \\
&&\hspace{5mm} -U_{tz} \hat{a}_{j\pm } \exp (\pm i\kappa _j x)]\bm{e}_y , 
\label{E_GI} \\
\hat{\bm{H}}_{j\pm } &=&-\frac{i}{\mu _0 } 
\sqrt{\frac{\hbar }{2\omega \varepsilon _0 V_j }} 
[U_{tz} ^\ast \hat{a}_{j\pm } ^\dagger \exp (\mp i\kappa _j x) \nonumber \\
&&\hspace{5mm} -U_{tz} \hat{a}_{j\pm } \exp (\pm i\kappa _j x)] 
(-\beta \bm{e}_x \pm \kappa \bm{e}_z ) \label{H_GI}
\end{eqnarray}
\end{subequations}
with double-sign in same order. 
Classical electromagnetic fields correspond to expectation values when in a coherent state.

States of optical fields in adjacent layers are related to each other 
by normalized Fresnel coefficients. Initial state $|\alpha \rangle _{j+} $ 
enters the interface between the $j$th and $(j+1)$th layers. 
After passing the interface, the state of the optical field becomes 
\begin{equation}
\hat{\mathcal{U}}_{j,(j+1)} |\alpha \rangle _{j+} |0\rangle _{(j+1)-} 
=|\tilde{t}_{j,(j+1)} \alpha \rangle _{(j+1)+} |-\tilde{r}_{j,(j+1)} \alpha \rangle _{j-} , 
\end{equation}
where 
\begin{equation}
\hat{\mathcal{U}}_{j,(j+1)} =\exp [\vartheta _{j,(j+1)} (\hat{a}_{j+} ^\dagger \hat{a}_{(j+1)-} 
-\hat{a}_{j+} \hat{a}_{(j+1)-} ^\dagger )]
\end{equation}
is a unitary operator and $\tan \vartheta _{j,(j+1)} =\tilde{r}_{j,(j+1)} 
/\tilde{t}_{j,(j+1)} $ (see Appendix \ref{BS_op}). 
We find that reflection disappears at the refractive index limit, 
varying continuously because 
\begin{eqnarray}
\tilde{r}_{j,(j+1)} \propto \kappa _j -\kappa _{j+1} 
=\frac{n_1 ^2 k_0 ^2 g^2 (2j+1)}{2\sqrt{n_1 ^2 k_0 ^2 -\beta ^2 } } \Delta x^2 \rightarrow 0
\end{eqnarray}
as $\Delta x\rightarrow 0$. Here an approximation 
\begin{equation}
\kappa _j \simeq \sqrt{n_1 ^2 k_0 ^2 -\beta ^2 } 
\left[ 1-\frac{n_1 ^2 k_0 ^2 g^2 }{2(n_1 ^2 k_0 ^2 -\beta ^2 )}x_j ^2 \right] 
\end{equation}
is considered. It means that the reflection of the optical field is negligible. 
Thus the optical field is totally transmitted with a refraction: 
$|\alpha \rangle _{j+} \rightarrow |\alpha \rangle _{(j+1)+} $. 
Similar procedures are repeated at the subsequent interfaces. 
The transmitted field travels according to Snell's law.

The optical field is totally reflected 
when the incident angle becomes larger than the critical angle $\theta _c $. 
This point is referred to as the turning point \cite{Snyder_text}. 
We assume that total reflection 
occurs when the optical field enters the layer where $j=t$: 
$|\alpha \rangle _{t+} \rightarrow |\alpha \rangle _{t-} $. 
Calculating expectation values of Eqs. (\ref{E_GI}) and (\ref{H_GI}) 
with a coherent state in each layer, we obtain electromagnetic fields 
that correspond to those in classical waveguide optics.

In the GI case, the optical field travels with a sinusoidal-like trajectory. 
This means that the optical field is not static in the cross-section, 
different from the SI case. An evanescent field is also generated at the turning points. 
However, it is small compared with the amplitude of trajectory 
of the optical field and is neglected here.

\begin{figure}[t]
\centering 
\includegraphics[width=.9\linewidth]{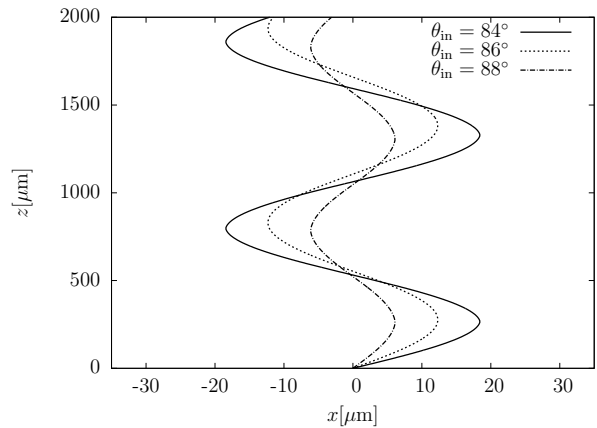}
\caption{\label{trajectory_q=2} Trajectories of optical fields in the GI waveguide. Here 
$q=2$ is used for the power-law profile in Eq. (\ref{power-law profile}). 
Parameters are assumed to be $\Delta x=0.1\ \mu \textrm{m}$, $n_1 =1.45$, 
and $g=5.66\ \textrm{m}^{-1} $. 
Trajectories of initial angles $\theta _{\rm in} =84^\circ, 86^\circ $, and 
$88^\circ $ are drawn as sinusoidal-like curves. }
\end{figure}

As a consequence, the optical field in the GI slab waveguide travels with a periodic trajectory. 
In Fig. \ref{trajectory_q=2}, trajectories of optical fields in the GI slab waveguide are shown. 
The initial state of the optical field is assumed to be $|\alpha \rangle _{0+} $, 
and the thickness of each layer $\Delta x=0.1\ \mu \textrm{m}$. 
Other parameters are as follows: 
$n_1 =1.45$ and $g=5.66\times 10^3 \textrm{m}^{-1} $; typical values 
for optical communication. Here the initial angles $\theta _{\rm in} =84^\circ ,
86^\circ $, and $88^\circ $ are studied. The results show that 
optical fields travel with sinusoidal-like trajectories 
with periods in phase with each other. These quantum optics results 
are consistent with classical waveguide optics \cite{Snyder_text}.

\section{Summary}
\label{summary}

We studied the propagation of an optical field in a dielectric from the perspective 
of quantum optics. 
We considered an interaction Hamiltonian between the optical field and an atom, 
and introduced the refractive index with the interaction Hamiltonian. 
We derived normalized Fresnel coefficients at an interface between different materials, 
and showed that they are related to the beam splitter operator. 
We also showed that, even for total reflection case, the evanescent field can exist 
outside of a reflection surface by a wavelength degree, 
with due consideration of the uncertainty relation.

We then studied the propagation properties of the optical fields in slab waveguides 
also using a viewpoint of quantum optics. 
A multi-layer division method was employed to represent the spatial dependence 
of the optical field in an inhomogeneous structure. 
For the step-index slab waveguide, 
we introduced spatially-averaged operators for static electromagnetic fields, 
and from them, we can evaluate the number of photons in each layer. The eigenvalue equation 
was derived in a similar way to classical waveguide optics. 
For the graded-index slab waveguide, we considered a power-law profile of the refractive index. 
We showed that optical fields travels with a sinusoidal-like trajectory, 
similar to those in classical waveguide optics.

When the coherent state shows high amplitude 
the results obtained here are reduced to those of classical waveguide optics. 
By contrast, for small amplitudes, the quantum properties of optical fields in waveguides 
can be determined with the present scheme.

\begin{acknowledgments}
The author would like to thank Prof. S.~Koh for valuable discussion. 
This work is supported by JSPS KAKENHI Grant Number 26790058. 
\end{acknowledgments}

\appendix 

\section{Derivation of Eq. (\ref{effective_Hamiltonian})}
\label{derivation_effective_Hamiltonian}

The interaction Hamiltonian between optical fields and an atom in a cavity 
was discussed in \cite{Schneider97}. A similar procedure is valid for the present case, 
and we review it in the following: 

We assume that the electric dipole approximation can be applied to the present model, 
and that the interaction Hamiltonian in the Schr\"{o}dinger picture is, under 
the rotating wave approximation, given by 
\begin{equation}
\hat{\mathcal{V}}=\sqrt{2}i\hbar g(\hat{A}\hat{\sigma }_+ -\hat{A}^\dagger 
\hat{\sigma }_- ), 
\end{equation}
where 
\begin{equation}
g=\sqrt{\frac{\omega }{2\hbar \varepsilon _0 V}}\bm{\wp }_{eg} \cdot \bm{e} 
\label{coupling_constant}
\end{equation}
is the coupling constant. Here $\bm{\wp }_{eg} $ and $\bm{e}$ 
are the electric dipole transition matrix element and a vector parallel 
to polarization direction, respectively. The operator 
\begin{equation}
\hat{A}=\frac{\hat{a}_p +\hat{a}_s }{\sqrt{2} } \label{A}
\end{equation}
and its Hermitian conjugate 
satisfy a commutation relation, 
\begin{equation}
[\hat{A},\hat{A}^\dagger ]=1. 
\end{equation}

In the interaction picture, we have 
\begin{eqnarray}
\hat{\mathcal{V}}_I (t)&=&\exp \left(i\frac{\hat{\mathcal{H}}_0 }{\hbar } t\right) 
\hat{\mathcal{V}}\exp \left(-i\frac{\hat{\mathcal{H}}_0 }{\hbar } t\right) \nonumber \\
&=&\sqrt{2}i\hbar g\left[ \hat{A}\hat{\sigma }_+ \exp (i\Delta \omega t)
-\hat{A}^\dagger \hat{\sigma }_- \exp (-i\Delta \omega t)\right] . \nonumber \\
&& \label{V_I}
\end{eqnarray}
The time-development operator is calculated as 
\begin{eqnarray}
\hat{U}_I (t)&\simeq &\hat{1}-\frac{i}{\hbar } \int _0 ^t dt' \hat{\mathcal{V}}_I (t' ) 
\nonumber \\
&&\hspace{10mm} -\frac{1}{\hbar ^2 } \int _0 ^t dt' \hat{\mathcal{V}}_I (t')
\int _0 ^{t' } dt'' \hat{\mathcal{V}}_I (t'' ) \nonumber \\
&\simeq &\hat{1}-2i\frac{g^2 }{\Delta \omega } [\hat{A}\hat{\sigma }_+ 
,\hat{A}^\dagger \hat{\sigma }_- ] t \nonumber \\
&\simeq &\exp \left( -i\frac{\hat{\mathcal{H}}' _{\rm eff}}{\hbar } t\right) , 
\end{eqnarray}
where 
\begin{equation}
\hat{\mathcal{H}}' _{\rm eff} =2\hbar \chi 
(\hat{\sigma }_+ \hat{\sigma }_- +\hat{A}^\dagger \hat{A}\sigma _z ) 
\label{H_eff'}
\end{equation}
and we introduce 
\begin{equation}
\chi =g^2 /\Delta \omega . \label{chi}
\end{equation}
In this calculation, 
the first integral is ignored as a small value under the assumption that 
the mean excitation $\langle \hat{A}\rangle 
\approx (\langle \hat{A}^\dagger \hat{A}\rangle )^{1/2}$ 
is not so large, that is, 
\begin{equation}
\left| \frac{g}{\Delta \omega } \sqrt{\hat{A}^\dagger \hat{A}}\right| \ll 1, 
\end{equation}
and terms proportional to $(\Delta \omega )^{-2} $ and any higher order is 
also ignored.

The atomic state remains in the ground state $|g\rangle _a $ before and after 
the interaction, and we calculate the expectation value of Eq. (\ref{H_eff'}) 
with respect to the atomic state. 
\begin{eqnarray}
_a \langle g|\hat{\mathcal{H}}_{\rm eff} ' |g\rangle _a 
&=&-\hbar \chi (\hat{a}_p \hat{a}_s ^\dagger +\hat{a}_p ^\dagger \hat{a}_s \nonumber \\
&&\hspace{10mm} +\hat{a}_p ^\dagger \hat{a}_p +\hat{a}_s ^\dagger \hat{a}_s ) , 
\label{<g|H_eff'|g>}
\end{eqnarray}
taking Eqs. (\ref{sigma_z|e>}), (\ref{sigma_z|g>}), (\ref{A}), and 
$\sigma _- |g\rangle _a =0$ into account. 
Since the third and fourth terms on the right hand side of 
Eq. (\ref{<g|H_eff'|g>}) are adequately small compared 
with the free Hamiltonian (\ref{free_Hamiltonian_field}) 
and they can be ignored, then we obtain 
the effective interaction Hamiltonian (\ref{effective_Hamiltonian}).

Assuming that the optical fields and atom interact during $\Delta \tau $, 
we have the time evolution from the initial state to the final one, 
\begin{equation}
_a \langle g|\hat{U}_I (\Delta \tau )|g\rangle _a 
=\exp \left( -i\frac{\hat{\mathcal{H}}_{\rm eff} }{\hbar } \Delta \tau \right) , 
\end{equation}
which appears in Eq. (\ref{final_state}).

\section{Beam splitter operator} \label{BS_op}

We introduce a unitary operator: 
\begin{equation}
\hat{\mathcal{U}}_{jj'}=\exp [i\vartheta (\hat{a}_j \hat{a}_{j' } ^\dagger 
+\hat{a}_j ^\dagger \hat{a}_{j' } )], \label{BS}
\end{equation}
where $\vartheta $ is a real parameter. 
Using this operator, we find that operators $\hat{a}_j $ and $\hat{a}_{j' } $ 
are transformed into \cite{Barnett_text} 
\begin{eqnarray}
\hat{\mathcal{U}}_{jj'} \hat{a}_j \hat{\mathcal{U}}_{jj'} ^\dagger 
&=&\hat{a}_j \cos \vartheta -i\hat{a}_{j' } \sin \vartheta , 
\label{BS_transform_1} \\
\hat{\mathcal{U}}_{jj'} \hat{a}_{j' } \hat{\mathcal{U}}_{jj'} ^\dagger 
&=&-i\hat{a}_j \sin \vartheta +\hat{a}_{j' } \cos \vartheta . 
\label{BS_transform_2} 
\end{eqnarray}
Two coherent states $|\alpha \rangle _j $ and $|\alpha ' \rangle _{j' } $ are combined by 
multiplying them by the operator (\ref{BS}) because 
\begin{eqnarray}
\lefteqn{\hat{\mathcal{U}}_{jj'} |\alpha \rangle _j |\alpha ' \rangle _{j' } } \nonumber \\
&=&\hat{\mathcal{U}}_{jj'} \hat{D}_j (\alpha )\hat{\mathcal{U}}_{jj'} ^\dagger 
\hat{\mathcal{U}}_{jj'} \hat{D}_{j' } (\alpha ' )\hat{\mathcal{U}}_{jj'} ^\dagger 
|0\rangle \nonumber \\
&=&\hat{D}_j (\alpha \cos \vartheta )\hat{D}_j (i\alpha ' \sin \vartheta ) \nonumber \\
&&\hspace{10mm} \times \hat{D}_{j' } (i\alpha \sin \vartheta )\hat{D}_{j' } 
(\alpha ' \cos \vartheta )|0\rangle \nonumber \\
&=&\hat{D}_j (\alpha \cos \vartheta +i\alpha ' \sin \vartheta ) \nonumber \\
&&\hspace{10mm} \times \hat{D}_{j' } (i\alpha \sin \vartheta +\alpha ' \cos \vartheta )
|0\rangle \nonumber \\
&=&|\alpha \cos \vartheta +i\alpha ' \sin \vartheta \rangle _j 
|i\alpha \sin \vartheta +\alpha ' \cos \vartheta \rangle _{j' } , \label{BS_transform_3}
\end{eqnarray}
where $\hat{D}_j (\alpha )=\exp (\alpha \hat{a}_j ^\dagger -\alpha ^\ast \hat{a}_j )$ is 
the displacement operator, and the relations (\ref{BS_transform_1}), (\ref{BS_transform_2}) 
and their Hermitian conjugates are considered. Here, 
the Baker-Hausdorff formula \cite{Louisell_text} 
and $\hat{\mathcal{U}}_{jj'} |0\rangle =|0\rangle $ are also considered.

By replacing $\hat{a}_{j' } $ with $-i\hat{a}_{j' } $, we obtain another unitary operator: 
\begin{equation}
\hat{\mathcal{U}}' _{jj'} =\exp [\vartheta (\hat{a}_j ^\dagger \hat{a}_{j' } 
-\hat{a}_j \hat{a}_{j' } ^\dagger )]. \label{BS'}
\end{equation}
The annihilation operators and coherent states are similarly transformed into 
\begin{eqnarray}
\hat{\mathcal{U}}' _{jj'} \hat{a}_j \hat{\mathcal{U}}^{\prime \dagger } _{jj'} 
&=&\hat{a}_j \cos \vartheta -\hat{a}_{j' } \sin \vartheta , 
\label{BS_transform_1'} \\
\hat{\mathcal{U}}' _{jj'} \hat{a}_{j' } \hat{\mathcal{U}}^{\prime \dagger } _{jj'} 
&=&\hat{a}_j \sin \vartheta +\hat{a}_{j' } \cos \vartheta . 
\label{BS_transform_2'} \\
\hat{\mathcal{U}}' _{jj'} |\alpha \rangle _j |\alpha ' \rangle _{j' } 
&=&|\alpha \cos \vartheta +\alpha ' \sin \vartheta \rangle _j \nonumber \\
&&\hspace{5mm} \otimes |-\alpha \sin \vartheta +\alpha ' \cos \vartheta \rangle _{j' } 
\label{BS_transform_3'}
\end{eqnarray}
with the operator (\ref{BS'}).

We can see from Eqs. (\ref{BS_transform_3}) or (\ref{BS_transform_3'}) 
that the optical fields of modes $j$ and $j' $ are combined 
with $\hat{\mathcal{U}}_{jj'} $ or $\hat{\mathcal{U}}'_{jj'} $. 
These operations physically correspond to a beam splitter.

\section{Derivation of Eqs. (\ref{ratio_coefficients}) and (\ref{eigenvalue_equation})}
\label{Appendix_C}
Here we derive relations betweeng eigenvalues of coherent states in areas (I), (II), and (III), 
and the eigenvalue equation from the continuity condition of the electromagnetic fields. 
Note that optical fields are totally reflected at interfaces between (I)-(II) and (II)-(III) 
and that there are evanescent fields in areas (I) and (III) in the step-index guiding case. 
The $x$-component of Poynting vector is calculated as a pure imaginary number, 
and it is not observable quantity. This means that energy flow from the core to cladding 
does not exist. 
Thus, normalization of the electromagnetic fields 
such as in Eq. (\ref{continuity_EH}) is not necessary.

For the TE even mode, for example, 
continuity conditions of the electromagnetic fields are 
\begin{subequations}
\label{continuity_even}
\begin{eqnarray}
\alpha ^{\rm (I)} \exp \left( -\frac{\gamma d}{2} \right) 
&=&\sqrt{2}\alpha ^{\rm (II)} _{\rm even} \cos \frac{\kappa d}{2} \nonumber \\
&=&\alpha ^{\rm (III)} \exp \left( -\frac{\gamma d}{2} \right) , \label{continuity_electric_even} \\
\gamma \alpha ^{\rm (I)} \exp \left( -\frac{\gamma d}{2} \right) 
&=&\sqrt{2}\kappa \alpha ^{\rm (II)} _{\rm even} \sin \frac{\kappa d}{2} \nonumber \\
&=&\gamma \alpha ^{\rm (III)} \exp \left( -\frac{\gamma d}{2} \right) . \label{continuity_magnetic_even}
\end{eqnarray}
\end{subequations}
From Eq. (\ref{continuity_electric_even}), eigenvalues are related 
as in Eq. (\ref{ratio_coefficients_even}). 
The condition (\ref{continuity_even}) is, in a matrix form, reduced to 
\begin{equation}
\left( 
\begin{array}{cc}
\exp \left( -\frac{\gamma d}{2} \right) & -\sqrt{2}\cos \frac{\kappa d}{2} \\
\gamma \exp \left( -\frac{\gamma d}{2} \right) & -\sqrt{2}\kappa \sin \frac{\kappa d}{2} 
\end{array}
\right) \left( 
\begin{array}{c}
\alpha ^{\rm (I)} \\
\alpha ^{\rm (II)} _{\rm even} 
\end{array}
\right) =0. 
\end{equation}
For a non-trivial solution, the determinant of the coefficient matrix should be zero, 
and we obtain the eigenvalue equation for the TE even mode in Eq. (\ref{eigenvalue_equation}). 
A similar procedure is applicable for the TE odd mode.

\bibliography{opt_fiber_q_refs}

\end{document}